# Linguistic Reflection in Java

GRAHAM KIRBY, RON MORRISON
*School of Mathematical and Computational Sciences, University of St Andrews, North Haugh,*
*St Andrews KY16 9SS, Scotland*
*email: {graham, ron}@dcs.st-and.ac.uk*

DAVID STEMPLE
*Department of Computer Science, University of Massachusetts, Amherst, MA 01003, USA*
*email: stemple@cs.umass.edu*

**SUMMARY**

**Reflective systems allow their own structures to be altered from within. Here we are concerned with a style of reflection, called linguistic reflection, which is the ability of a running program to generate new program fragments and to integrate these into its own execution. In particular we describe how this kind of reflection may be provided in the compiler-based, strongly typed object-oriented programming language Java. The advantages of the programming technique include attaining high levels of genericity and accommodating system evolution. These advantages are illustrated by an example taken from persistent programming which shows how linguistic reflection allows functionality (program code) to be generated on demand (Just-In-Time) from a generic specification and integrated into the evolving running program. The technique is evaluated against alternative implementation approaches with respect to efficiency, safety and ease of use.**

## INTRODUCTION

Linguistic reflection may be defined as the ability of a running program to generate new program fragments and to integrate these into its own execution. The focus of this paper is to show how this style of reflection may be provided in the compiled, strongly typed language Java and used as a paradigm for program generation. In particular, we describe how linguistic reflection may be used to build code *Just-In-Time* from a generic specification.

Two techniques for type-safe linguistic reflection have evolved: compile-time linguistic reflection[1] and run-time linguistic reflection[2, 3]. In this paper we will only consider run-time linguistic reflection, in which the executing program generates new program fragments in the form of source code, compiles these fragments using a dynamically callable compiler, and finally links the compiled fragments into its own execution. The process may be recursive, since the generated source fragments may contain further reflective code that only reflects when it is executed. The technique should be clearly distinguished from the style of reflection supported by Metaobject Protocol systems[4, 5], in that our emphasis is on the type-safe generation and binding of new code. Linguistic reflection has also been termed structural reflection[6].

The ability of a linguistically reflective program to significantly change its behaviour dynamically by adding new code reduces the opportunities for static type checking. It is therefore not surprising to find a tension between linguistic reflection and type checking. The trick is to provide the linguistic reflective facilities in a manner that retains as much static type checking as possible without being so severe as to remove all the benefits of reflection. This may be done by providing a meta-level description of the types of objects which can be used in the generation of new program fragments. In run-time linguistic reflection, once the fragments have been generated, type checking occurs in both the compilation of the new fragments and their linking into the running program.

The motivation for this work comes from the desire for two advanced programming capabilities. The first is the ability to implement highly abstract (generic) specifications, using a meta-level description of types, within a strongly typed programming language. The second is the ability to accommodate some of the continual changes (evolution) in data-intensive applications without resorting to highly interpretative approaches or *ad hoc* restructuring methods. Both capabilities involve reflective access to the types (classes) of a system that is changing itself.

These motivations are particularly relevant in persistent systems since such systems allow data to retain its type structure throughout its lifetime. Our own experience brought to the Java programming context is with strongly typed run-time linguistic reflection in the persistent languages PS-algol[7] and Napier88[8]. So although the work is described here in the context of standard Java, it has been applied to a persistent version of Java, PJama[9], for maximum effect.

We will emphasise generation of programs here, but it should be noted that the technique has been extensively used to control system evolution in a type safe manner. An application, database or persistent system needing to evolve may change its code and operation using linguistic reflection.

The motivating example, to write a strongly typed generic natural join function, is taken from persistent programming. The subtlety of the example is that the function cannot be exactly typed by conventional polymorphic type systems since the code for the function and its result type depend upon the types of the parameter values. The difficulty of typing such a generic function can be avoided by using linguistic reflection to generate specific, simply typed, functions on demand[*]. Once generated the functions are compiled and bound to the executing program and applied, hence the use of the term *Just-In-Time* generation.

# RUN-TIME LINGUISTIC REFLECTION

In run-time linguistic reflection, the executing program generates new program fragments in the form of source code, invokes a dynamically callable compiler, and finally links the results of the compilation into its own execution. The generation of the source code is most effective where the values and types of existing data are used to tailor the generated code to that data. This requires three distinct facilities:
- a meta-level mechanism that provides descriptions of the types of existing data;
- a compiler that can be invoked dynamically; and
- a dynamic incremental loader.

We will show in the next section how these facilities can be provided in Java. First we sketch a motivating example for linguistic reflection, which will be explored in detail later in the paper.

**Overview of a Motivating Example Application**

A generic relational natural join function provides our motivating example. The function allows a join operation on any pair of relations. Without such a generic function a separate join function would have to be written for every pair of relation types. The major benefit of genericity is therefore brevity of specification. The drawback is that most statically typed systems cannot accommodate such specifications. In these cases linguistic reflection may be used to provide a solution.

The generic natural join function represents an abstraction over types that is beyond the capabilities of even the most polymorphic type systems. The details of the input types, particularly the names and domains of the attributes, significantly affect the algorithm and the output type of the function, determining:
- the result type;
- the code to test whether a given pair of tuples matches on the overlapping attributes; and
- the code to build a relation having tuples with the aggregation of attributes from both input relations but with only one copy of the overlapping attributes.

The type of a polymorphic natural join function would be:

$$\forall a.\forall b.(\text{relation } [a] \times \text{relation } [b] \rightarrow \exists c.\text{relation } [c])$$
where $c$ is some type related to $a$ and $b$ by the rules of natural join

That is, the function takes as parameters two relations, with tuple types $a$ and $b$ respectively, and returns a third relation with a related tuple type $c$ as a result. Type systems that can accommodate such functions are generally higher order and dynamically checked, since the relationship between the input types (which are unknown statically) and the output type precludes static checking. An interesting exception in this case is the Machiavelli type system[10]. In general, however, in order to retain static checking and still write such higher-order functions, a novel technique is required. One such technique is linguistic reflection.

The function cannot be written as a statically typed polymorphic procedure since it requires knowledge of the type structure of $a$, $b$ and $c$ and they are explicitly abstracted over here. The linguistic reflection technique allows a meta-function to be written, the generic natural join, that can examine the structure of the input types. While this generic meta-function cannot be statically typed in general, for any one particular call of the generic meta-function the type structures *are* known. At the point of use of the natural join, the generic meta-function is called and given as input the type structure. It then computes a specific natural join function for these input types together with a call to the specific natural join. This generated code is then fed into the compiler dynamically for checking and code generation, then linked and executed.

---

[*] It may be thought at first that static discovery of the input types would allow generation techniques such as Ada generics to be used. However, Ada generics do not allow computation over the type representations, are not recursive, and do not generate *Just-In-Time*.

The generic natural join meta-function involves computing over the representations of the input types, thus the language must provide facilities for this. Indeed it is this ability that contrasts the reflective language with other polymorphic systems.

In general, when using run-time linguistic reflection it is possible to generate code based on the inspection of values not just types. For example, in the natural join case it is possible to discover the input types dynamically and to examine the size of the input relations in order to generate an optimised algorithm. Hence a second reason for the motivating example. Such generation and possible optimisation is type-safe and performed before execution, but beyond the scope of static type systems.

**Other Example Applications**

Two other examples of using linguistic reflection to achieve type-safe genericity are a generic *map* function, and an object browser[†]. The first example involves writing a function which takes as parameters a data structure of any type, and a function, and returns a copy of the data structure in which every occurrence of the parameter function's input type is replaced by the result of applying the parameter function. This is so general that it cannot be simply typed. A specialised map function for lists, for example, could be typed as:

$$\forall a.\forall b.(list[a] \times (a \rightarrow b) \rightarrow list[b])$$

This applies a function of type $(a \rightarrow b)$ to every element of the input list to form the new list. Thus the input and output types of the function parameter are abstracted over. There is however a second possible abstraction, where the data structure type itself becomes a parameter rather than being specifically the list constructor. Like the natural join example, this cannot be expressed as a statically-typed polymorphic function because of the complex relationship between the input and output types, and because the algorithm to build the result depends on the type structure of the first parameter. These problems can be overcome by using linguistic reflection to generate, compile and link appropriate code as needed, based on examination of the inputs.

The second example, the object browser, involves writing a generic function that can display a graphical representation of any value passed to it. The type of the browser function would be, using existential type notation[12]:

$$(\exists a.a)$$

Thus the browser function has a statically known type, being a function that takes a value of some type about which nothing is known[‡]. This example is slightly different from the first in that there is no difficulty in typing the function in a conventional polymorphic system, but the algorithm used to produce the graphical display depends on the type structure of the input value. Whereas a polymorphic function abstracts over the parameter type, the browser needs details of that type in order to produce the display. Again, this can be achieved with a linguistic reflective solution which generates, compiles and links new code based on the input type.

# SUPPORTING STRONGLY TYPED RUN-TIME LINGUISTIC REFLECTION IN JAVA

We now show how each of the necessary facilities for linguistic reflection—meta-level descriptions, dynamic compilation and incremental loading—is provided in Java. The following section will illustrate how they are combined with generator programs to achieve type-safe linguistic reflection.

**Meta-level Description of Classes**

The first requirement for run-time linguistic reflection is the availability of meta-level descriptions of type information. These are provided in Java through the core reflection package and in particular the class *Class*. Each Java class has an associated unique instance of this class[§]. A class representation may be obtained by calling the following method, inherited by every object from the root class *Object*:

---

[†] More details of these examples can be found in the papers in which they were originally described, by Sheard[1], and Dearle & Brown[11] respectively.

[‡] For example type *dynamic* in Amber[13] or type *any* in Napier88[8].

[§] All arrays of a particular dimension and element class share a single class object.

```
    public final Class getClass();
```

The core reflection package *java.lang.reflect* contains the class *Class*, together with the associated classes *Array*, *Constructor*, *Field*, *Method* and *Modifier*. Part of the definition of *Class*, that relevant to the discussion, is shown in Figure 1; details of exceptions have been omitted to save space.

```
1   public final class Class {
2
3       public static Class forName( String className );
4       public Object newInstance();
5       public String getName();
6       public Class getSuperclass();
7       public Class[] getInterfaces();
8       public int getModifiers();
9       public Class[] getDeclaredClasses();
10      public Field[] getDeclaredFields();
11      public Method[] getDeclaredMethods();
12      public Constructor[] getDeclaredConstructors();
13      public Field getDeclaredField( String name );
14      public Method getDeclaredMethod( String name, Class parameterTypes[]);
15      public Constructor getDeclaredConstructor( Class parameterTypes[] );
16      …
17  }
```

*Figure 1. The class* Class

An instance of *Class* can be manipulated by a running Java program. The user cannot create such instances since *Class* has no public constructors; instead they are constructed by the Java Virtual Machine (JVM) when classes are loaded. As an example, Figure 2 illustrates how a program fragment can print the names of the methods of the class in which the fragment occurs.

```
1   Class thisClass = this.getClass();
2   Method[] methodsOfThisClass = thisClass.getDeclaredMethods();
3   for ( int i=0; i < methodsOfThisClass.length; i++ ) {
4       System.out.println( methodsOfThisClass[i].getName() ); }
```

*Figure 2. Example use of meta-level information*

The availability of information on the structure of classes via *Class* instances allows programs to generate new programs based on analysis of existing objects. A complete description of the class of an object may be built up by invoking the object's *getClass* method, followed by invoking the various *Class* methods to obtain representations of the class's fields, methods, constructors, superclass and interfaces. These representations in turn provide methods that allow their structure, for example the types of fields and the parameter types of methods, to be discovered. Where such components are object types the process can be continued recursively.

Given the ability to discover the structure of a class, it is possible to write generic methods that generate new code fragments that exploit this information, such as the examples described in the previous section.

**Incremental Loading of Program Fragments**

The second and third requirements for run-time linguistic reflection are dynamic compilation and the ability to load new code dynamically. For ease of explanation we discuss the latter first.

Java supports the ability to load new code through the dynamic loading of classes. This is achieved by defining and using a subclass of the abstract class *ClassLoader*, the signature of which is partially defined in Figure 3.

```
1   public abstract class ClassLoader {
2
3       public Class            loadClass( String name );
4       protected abstract Class loadClass( String name, boolean resolve );
5       protected final Class   defineClass( byte data[], int offset, int length );
6       protected final void    resolveClass( Class c );
7       protected final Class   findSystemClass( String name );
8       …
9   }
```

*Figure 3. The class* ClassLoader

Note that *defineClass*, *resolveClass* and *findSystemClass* are **final** and therefore may not be overridden in a subclass, while *loadClass (String, boolean)* is **abstract** and must be overridden. The main task in defining a new class loader is to specify how a given class name is mapped to a corresponding byte sequence containing the compiled class details. Figure 4 shows an example class loader that attempts to load class definitions from a specified file system directory.

```
1   import java.io.*;
2
3   public class DirSpecificClassLoader extends ClassLoader {
4
5       private File classDirectory;
6
7       DirSpecificClassLoader( File classDirectory ) {
8           this.classDirectory = classDirectory; }
9
10      protected Class loadClass( String name, boolean resolve )
11          throws ClassNotFoundException {
12
13      Class resultClass;
14
15      try { resultClass = findSystemClass( name ); } catch (Exception e1) {
16
17          try {
18              File classFile = new File( classDirectory, name + ".class" );
19              if ( ! classFile.exists() ) throw new ClassNotFoundException();
20
21              FileInputStream classFileInputStream =
22                  new FileInputStream( classFile );
23
24              int fileLength = classFileInputStream.available();
25              byte[] classBytes = new byte[ fileLength ];
26              classFileInputStream.read( classBytes );
27
28              resultClass = defineClass( classBytes, 0, classBytes.length );
29          }
30
31          catch ( IOException e2 ) { throw new ClassNotFoundException(
32              "reading class file: " + e2.getMessage() ); }
33          catch ( ClassFormatError e3 ) { throw new ClassNotFoundException(
34              "class file format: " + e3.getMessage() ); }
35      }
36
37      if ( resolve ) resolveClass( resultClass );
38      return resultClass;
39      }
40  }
```

*Figure 4. A* ClassLoader *that loads from a specified directory*

In line 8 the constructor records the directory from which class definitions will be loaded. The only method to be overridden is *loadClass (String, boolean)*. The method first attempts to return a previously loaded version of the class by calling the inherited method *findSystemClass*. If the class has not already been loaded, this will cause an exception, which is ignored. The method then tries to access a *.class* file in the directory specified at instantiation. If this fails, a *ClassNotFoundException*. is thrown.

Assuming that the appropriate *.class* file is located in the directory, lines 21-26 create a byte array and read the contents of the file into the array. An *IOException* may be thrown at this point. The inherited method *defineClass* is then called in line 28 to coerce the byte array into a new *Class* instance. This checks whether the bytes define a valid Java class, using any byte-code verification in use by the JVM, and throws a *ClassFormatError* if not. Finally, if specified by the parameter *resolve*, references to other classes are resolved by calling the inherited method *resolveClass*, which itself calls *loadClass* for any unloaded classes referenced by the class.

Once a class has been loaded at run-time, an instance of the class can be created as illustrated in Figure 5 and Figure 6. Note that it is initially typed as an instance of the root class *Object*, since the *newInstance* method returns *Object*. If any class-specific methods of the new instance are to be invoked, their signatures must be known statically and the instance cast to a class or interface defining them. In the example, the new instance of the dynamically loaded class *Person* is cast to the statically known interface *PersonInterface*.

```
1   File classDir = new File( "/user/graham/classes" );
2   DirSpecificClassLoader loader = new DirSpecificClassLoader( classDir );
3   try {
4       Class newClass = loader.loadClass( "Person" );
5       Object aPerson = newClass.newInstance();
6
7       // Assumes class Person was defined to implement PersonInterface.
8       PersonInterface usablePerson = (PersonInterface) aPerson;
9       usablePerson.writeAge();
10  } catch ( Exception e ) { System.out.println( "class loading error" ); }
```

*Figure 5. Using a dynamically loaded class*

```
1   interface PersonInterface {
2       public void writeAge();
3   }
```

*Figure 6. A statically known interface*

This demonstrates that the ability to dynamically load new code, required for linguistic reflection, is indeed a feature of Java.

**Dynamic Compilation**

Given the ability to load classes dynamically, it is possible to define a class to support dynamic compilation, the final ingredient necessary for integrating reflectively generated code into running programs. We will explore three techniques for providing dynamic compilation in Java:
- The first writes the source code to the file system and then issues an operating system command to invoke the compiler on the new source files. The compiler creates a corresponding set of class files that are then loaded into the running program.
- The second is a variation on the first which exploits the existence of a compiler written in Java and available through class libraries. This avoids the need to invoke the compiler via an operating system command, since it can be called directly.
- The third does not access the file system at all. Instead, it uses a compiler written in Java, which takes a set of source class representations as a parameter, and returns a set of byte arrays containing the compiled byte code, from which the classes may be loaded.

Regardless of which technique is used to implement dynamic compilation, the same class interface can be presented to the programmer. Figure 7 shows an outline of such a class. The method *compileClasses* takes a pair of string arrays containing the names and source definitions of the classes to be compiled, and attempts to compile them to an array containing instances of class *Class*[**]. More complex source representations such as abstract syntax trees could be accommodated by overloading the *compileClasses* method.

```
1   public class Compiler {
2       // A run-time callable compiler.
3       public Class[] compileClasses( String[] classNames, String[] classDefns )
4           throws CompilationException {…}
5       …
6   }
7
8   public class CompilationException extends Exception {
9       // Represents a compilation error arising from an invalid program
10      // or some other error occurring during compilation.
11      …
12  }
```

*Figure 7. Outline of a compiler class*

The first implementation mechanism involves writing the source code to a set of files, reading the source files, writing out a further set of compiled *.class* files, reading the *.class* files, and finally converting them to *Class* instances using a class loader such as was shown in Figure 4. This is done because standard Java compilers work with source and *.class* files, thus such compilers can be used without change. The compiler is invoked by creating a new operating system process to call the Java compiler. The technique has two disadvantages, however:
- Details of how to invoke the compiler vary depending on which platform Java is running on. On some platforms it is not even possible to create new operating system processes from a Java program.
- Some Java compilers—such as the Sun Microsystems release compiler at the time of writing—are implemented in Java itself. Invoking such a compiler in a new operating system process thus involves starting a new instantiation of the JVM, consuming significant additional run-time resources.

---
[**] Although the class names could be extracted by parsing the source definitions, here they are passed as parameters to avoid the need for parsing.

The mechanism is illustrated in Figure 8, which shows a private *compile* method, tailored for a Unix environment, that could be used by the public *compileClasses* method. In lines 12-13 it reads various environment settings. Then, for each class source definition, it creates a *.java* file in the compiler directory and writes the definition to it. Lines 24-26 construct a Unix shell command, using the appropriate classpath and output directory, to invoke the compiler on all the *.java* files in the directory; the command is then executed as a new process in lines 28-30, while the current program blocks waiting on the result. The result code returned by the shell command is used to determine whether the compiler was invoked successfully. If so, the method attempts to read the new class definitions from the compiled class files. This will fail if a compilation error has occurred due to an invalid program, since the class files will not have been generated.

For this to work more generally, rather than just on Unix, the Java host environment must be interrogated in order to deduce the appropriate compiler invocation command, and indeed whether or not it is possible to issue an operating system command. In some environments the technique does not work at all, for example when the host operating system has no command line interface, or when running as an applet within a web browser which disallows operating system access.

```
1   import java.io.*;
2
3   public class Compiler {
4       …
5       private String compilerPath = …;    // Path of compiler executable.
6       private String compilerDirPath = …; // Path of compiler directory.
7
8       private Class[] compile( String[] classNames, String[] classDefns )
9           throws CompilationException {
10
11      try {
12          String javaClassPath = System.getProperty( "java.class.path" );
13          String fileSeparator = System.getProperty( "file.separator" );
14
15          for (int i = 0; i < classNames.length; i++) {
16
17              File compilerDir = new File( compilerDirPath );
18              File sf = new File( compilerDir, classNames[i] + ".java" );
19
20              PrintWriter p = new PrintWriter( new FileOutputStream(sf) );
21              p.println( classDefns[i] );
22          }
23
24          String compileCommand = compilerPath + " -classpath " + javaClassPath +
25                                  " -d " + compilerDirPath + " " +
26                                  compilerDirPath + fileSeparator + "*.java";
27
28          Process cp = Runtime.getRuntime().exec( compileCommand );
29          if ( cp.waitFor() != 0 )
30              throw new CompilationException( "invocation of compiler failed" );
31
32          // Load classes from .class files using class loader, and return as array.
33          // If class files can't be found, throw CompilationException.
34          …
35      }
36      catch ( Exception e ) { throw new CompilationException( "compilation error" ); }
37      }
38  }
```

*Figure 8. Details of private* compile *method*

The second implementation technique exploits the availability of a Java compiler written in Java and accessible through a class library, where such a compiler exists, executing the compiler directly on the existing instantiation of the JVM. This reduces run-time overheads and avoids the need for the issue of commands to the operating system, since the compiler can be loaded and invoked like any other Java class.

Figure 9 shows the code with which lines 24-30 of Figure 8 would be replaced. The string array *argv* contains the various compilation parameters, specifying the classpath, where the output class files are to be created, and which source files are to be compiled (in this case all the *.java* files in a temporary directory). A new instance of the compiler class *Main* is created, specifying that error messages are to be written to the standard output. Finally the *compile* method is invoked and a result code returned.

The disadvantage of this technique is the reliance on knowledge of the Java implementation, in particular of the compiler API and of which package contains the compiler. Thus a change in the Java implementation—such as placing the compiler in a different package or re-implementation of the compiler in a different language—would lead to compilation failure.

```
1    // Initialise strings 'javaClassPath', 'tempDirPath' and 'fileSeparator'.
2
3    String[] argv = { "-classpath", javaClassPath, "-d", tempDirPath,
4                      (tempDirPath + fileSeparator + "*.java") };
5
6    sun.tools.javac.Main compiler = new sun.tools.javac.Main( System.out, "javac" );
7
8    if ( ! compiler.compile( argv ) )
9        throw new CompilationException( "invocation of compiler failed" );
```

*Figure 9. Direct invocation of Java compiler*

Both variations described so far involve a sequence of writing source code to the file system, reading it again, writing compiled code, and finally reading in the compiled code. For the purposes of dynamic compilation the versions written to the file system are no longer needed once the compiled *Class* instances have been obtained, and so the file access represents undesirable disk I/O. Furthermore in some environments, for example web browsers, the JVM may be executing in the context of a security manager which disallows access to the local file system. Thus it would be advantageous to avoid file access during dynamic compilation.

The third implementation technique involves a refinement of the compiler API to allow a source definition to be compiled directly to a set of byte arrays, from which the classes can be loaded without access to the file system[††]. This requires access to the source code of a compiler written in Java. An additional method *compileToBytes* is required:

```
1    public class Compiler {
2        …
3        private byte[][] compileToBytes( String[] classNames, String[] classDefns )
4            throws CompilationException { … }
5    }
```

*Figure 10. Outline of private* compileToBytes *method*

The public *compileClasses* method now calls the *compileToBytes* method to obtain the compiled code directly without file access, in the form of an array of byte arrays, each containing the code for one class. It then uses an instance of *ByteArrayClassLoader* to load the classes from the byte arrays, as defined in Figure 11. The constructor takes as parameter a particular byte array which is stored in the *compiledClass* field. The *loadClass* method then reads bytes directly from that array.

---

[††] Whereas the authors have implemented the first two dynamic compilation mechanisms, this third mechanism has not yet been implemented due to the changes to the compiler source code that would be required.

```
1   class ByteArrayClassLoader extends ClassLoader {
2
3       private byte[] compiledClass;
4
5       ByteArrayClassLoader( byte[] compiledClass ) {
6           this.compiledClass = compiledClass; }
7
8       protected Class loadClass( String name, boolean resolve )
9           throws ClassNotFoundException {
10
11          try {
12              Class resultClass = defineClass( compiledClass, 0, compiledClass.length );
13
14              if ( resolve ) resolveClass( resultClass );
15              return resultClass;
16          }
17
18          catch ( ClassFormatError e ) { throw new ClassNotFoundException(
19              "class file format: " + e.getMessage() ); }
20      }
21  }
```

*Figure 11. A* ClassLoader *that loads from a byte array*

**Summary**

The requirements for run-time linguistic reflection were earlier identified as the provision of meta-level class information, the ability to load new classes dynamically, and the ability to call the compiler dynamically. This section has described how a range of implementation techniques may be used to take a source-level program representation and convert it to an array of *Class* objects at run-time. Instances of these classes may be created dynamically using the *newInstance* method. The methods of the newly created instances are then available to the running application, completing the process of linguistic reflection. The next section illustrates a use of this ability to generate and use new code dynamically.

# THE MOTIVATING EXAMPLE

**Implementation of Generic Natural Join Using Linguistic Reflection**

We now describe how the motivating example of generic natural join can be implemented in Java. Space considerations preclude a complete description of the linguistic reflective solution here[‡‡]. We give an overview, consisting of example relation classes and calls of the generic natural join function (Figure 12 and Figure 13); part of the reflective implementation of the generic natural join function (Figure 15 and Figure 16), and code generated by an example call (Figure 17).
  The implementation involves two main parts:
- Selecting a representation of relations as Java classes. There is a choice between using a single class which is sufficiently general to represent all relational types, or using a different class for each different type. The second option is chosen here.
- Defining a class with a generic static method to perform the natural join operation on any pair of relations, using *Just-In-Time* generation of specific source code. Since the method accepts any relations, its two formal parameters and its result are typed as the general class *Object*, while its actual parameters on a call will have classes representing specific relational types.

  The technique chosen for representing relations is to define a class for each relational tuple type, specifying the attributes of that type. Each relation is then represented as an array of tuples, each of which is an instance of the class. Although an ordering on the tuples is introduced which is not part of the relational semantics, this can be ignored.
  The simplest representation would be to define, for each attribute, a public field in the relation class. However, as will be explained shortly, the natural join solution requires the use of Java interfaces, as well as classes, to represent relational types. Since interfaces cannot define public fields, public methods rather than fields are used to represent attributes. Each relational tuple class or interface thus contains one parameter-less method for each attribute, which returns the corresponding attribute value.

---

[‡‡] The full Java source code for the example is available at:
    http://www-ppg.dcs.st-and.ac.uk/Java/ReflectionExample/

The representation is illustrated in Figure 12, in which lines 1-20 define the class *Employee*. The four public methods *name()*, *title()*, *department()* and *jobId()* correspond to the attributes of the relation, and each relation instance is represented as an array of type *Employee[]*. The classes *Job* and *Salary* are defined similarly.

```
1   public class Employee {
2
3       private String  name;
4       private String  title;
5       private String  department;
6       private int     jobId;
7
8       public String name()        { return name; }
9       public String title()       { return title; }
10      public String department()  { return department; }
11      public int    jobId()       { return jobId; }
12
13      public Employee( String name, String title, String department, int jobId ) {
14
15          this.name =        name;
16          this.title =       title;
17          this.department =  department;
18          this.jobId =       jobId;
19      }
20  }
21
22  // Attributes 'post', 'duties' and 'jobId'. Details omitted to save space.
23  public class Job {…}
24
25  // Attributes 'post' and 'salary'. Details omitted to save space.
26  public class Salary {…}
```

*Figure 12. Classes representing relational tuple types*

Figure 13 shows two invocations of the generic natural join method. The relations are formed in lines 2-4, calling utility methods to read data from files into arrays of the appropriate types. An instance of the generic natural join class *NatJoin* is then created. In line 8 a call is made to the method *natJoin* which performs the natural join on the relations *emps* and *jobs*. The specific type of the resulting relation *employeesAndJobs* is not known statically, thus it is typed as *Object*. The contents of the relation are printed out by a call to the method *printRelation* in line 9; this method is implemented in a similar way to *natJoin* in that it inspects the type of the relation passed to it and then generates specific code to print out relations of that type. Finally, in lines 11-12 the result relation is joined with a third relation *salaries*, and the result of that join is printed out. Note that since the input and output types of *natJoin* are *Object*, the same method is used to perform both example joins that have different actual parametric types. Also, in this example the result types of the joins need not be specified statically.

```
1   try {
2       Employee[]  emps =     loadEmployees( "employees.txt" );
3       Job[]       jobs =     loadJobs( "jobs.txt" );
4       Salary[]    salaries = loadSalaries( "salaries.txt" );
5
6       NatJoin natJoinPack = new NatJoin();
7
8       Object employeesAndJobs = natJoinPack.natJoin( emps, jobs );
9       natJoinPack.printRelation( employeesAndJobs );
10
11      Object empsJobsAndSalaries = natJoinPack.natJoin( employeesAndJobs, salaries );
12      natJoinPack.printRelation( empsJobsAndSalaries );
13  }
14  catch (Exception e) { System.out.println( "join failed: " + e.getMessage() ); }
```

*Figure 13. Invoking linguistic reflective Natural Join*

In the previous example there is no need to specify the result types of the joins in the program. This may be an advantage in that there is less for the programmer to write down. However, the drawback is that since the result type is not specified statically, it is not possible for the program to perform any operation on the result relation which requires static knowledge of its specific type—for example, accessing the value of a particular tuple attribute.

To accommodate such uses of the result relation, the natural join class provides an alternative *natJoin* method that takes an interface representation as an additional parameter. The method operates in the same way as the original *natJoin* except that the class representing the result tuples is defined to implement the specified interface. This means that it is possible for the calling program to cast the result relation to the interface, making the relation available with a specific type.

A call of the second *natJoin* method is shown in Figure 14. The definitions of *Employee* and *Job* are as before; the interface *EmpJob* specifies the expected result type, containing the union set of the methods from *Employee* and *Job*. The loading of the relations *emps* and *jobs* is as before and is omitted here. Line 16 calls the static method *forName* from class *Class* to obtain a representation of *EmpJob*, which is passed as the third parameter to *natJoin*. The result relation is then cast to the specific type *EmpJob[]* for use in the remaining part of the program.

```
1    public class Employee {…}
2    public class Job {…}
3
4    public interface EmpJob {
5
6        public String name();
7        public String title();
8        public String department();
9        public int    jobId();
10       public String post();
11       public String duties();
12   }
13
14   try {
15       …
16       Class resInterface = Class.forName( "EmpJob" );
17       EmpJob[] result = (EmpJob[]) natJoinPack.natJoin( emps, jobs, resInterface );
18       …    // 'result' now available with specific type.
19   }
20   catch (Exception e) { System.out.println( "join failed: " + e.getMessage() ); }
```

*Figure 14. Specifying a statically defined result type*

The previous figures showed how the generic natural join methods are used. The signature of the main *natJoin* method is shown in line 3 of Figure 15: it takes two relations, typed as *Object*, and returns the result of the natural join as an *Object*. This method simply calls the alternative *natJoin* method, passing *null* as the additional interface parameter.

Lines 10-16 extract the types of the input relations using the core reflective methods *getClass* and *getComponentType*, checking that the parameters are valid arrays. Lines 19-28 generate names and source code for the new classes. Two new class source representations are generated: the main one containing a method to perform the natural join for the current pair of input types, and another auxiliary class defining the result type. If a non-null expected result interface is supplied as the parameter *resInterface*, the result class is defined as implementing that interface. The main generation is performed by the methods *generateJoin* and *generateResClass*, which take as parameters the class names, input types and various other information which is omitted here for brevity.

Once the source code strings for the two classes have been generated, they are compiled and loaded by the method *compileAndReturnMainClass* which compiles all the classes passed to it and returns the *Class* instance corresponding to the specified principal class. In this case the principal class is the one which defines the generated natural join method *join*. That method is obtained in line 37 with a call to *getMethod*, passing it the name "join" and an array of *Class* instances specifying the parameter types (two *Object* parameters). Finally, the join method is invoked on the input relations, using the method *invoke*.

```
1    // 'Linguistic reflective' implementation of natural join.
2    public class NatJoin {
3        public Object natJoin( Object rel1, Object rel2 ) throws InvalidJoinException {
4            return natJoin( rel1, rel2, null );
5        }
6
7        public Object natJoin( Object rel1, Object rel2, Class resInterface )
8            throws InvalidJoinException {
9
10           // Get classes representing the tuples of the relations to be joined.
```

```
11          Class tuple1Class = rel1.getClass().getComponentType();
12          Class tuple2Class = rel2.getClass().getComponentType();
13
14          // Check the input relation objects are arrays.
15          if ( tuple1Class == null || tuple2Class == null )
16              throw new InvalidJoinException( "Invalid input relations" );
17
18          try {
19              // Get names for the new classes to be generated.
20              String[] classNames = { uniqueId(), uniqueId() };
21              String[] classDefns = new String[ 2 ];
22
23              …
24
25              // Generate the join algorithm and the result class definition.
26              classDefns[ 0 ] = generateJoin( classNames[ 0 ], classNames[ 1 ],
27                                              tuple1Class, tuple2Class, … );
28              classDefns[ 1 ] = generateResClass( classNames[ 1 ], resInterface, … );
29
30              // Compile the generated class definitions.
31              Class mainGeneratedClass = compileAndReturnMainClass(
32                  classNames, classDefns, classNames[ 0 ] );
33
34              // Get the 'join (Object,Object)' method.
35              Class objectClass = (new Object()).getClass();
36              Class[] joinArgClasses = { objectClass , objectClass };
37              Method join = mainGeneratedClass.getMethod( "join", joinArgClasses );
38
39              // Invoke the 'join' method on the relations.
40              Object[] joinArgs = { rel1, rel2 };
41              return join.invoke( null, joinArgs );
42          }
43          catch (Exception e)
44              { throw new InvalidJoinException( "join failed: " + e.getMessage() ); }
45      }
46
47      private String generateJoin(…}
48      private String generateResClass( String className, Class resInterface, … ) {…}
49      private String uniqueId() {…}
50 }
```

*Figure 15. Linguistic reflective implementation of Natural Join*

Figure 16 shows some of the details of the main generator method *generateJoin*. Its parameters are the names of the classes to be generated, the input and output types for the join, and arrays containing the intersection and union of the attributes (represented as the corresponding methods) of the two input relations. These are computed in the omitted part of the *natJoin* method in Figure 15. Line 7 creates a new program representation, an instance of the class *Program* which manipulates string source representations and provides methods for adding source fragments, controlling indentation, etc. Lines 9-18 generate *import* statements, explanatory comments and the class header; lines 22-23 generate the header of the join method. Lines 27-51 generate the *match* method; the remainder of the generation code is omitted for brevity.

```
1  private String generateJoin(
2      String joinClassName, String resultClassName,
3      Class tuple1Class, Class tuple2Class, Class resInterface,
4      Method[] commonAttributes, Method[] allAttributes ) {
5
6      // Create a new program representation.
7      Program p = new Program();
8
9      // Generate the necessary import statements and class header.
10     p.addLn( "import java.util.*;" );
11     p.addLn();
12     p.add( "// This is a generated class to join " + tuple1Class.getName() );
13     p.addLn( "s and " + tuple2Class.getName() + "s." );
14     if ( resInterface != null ) {
15         p.add( "// The result of method join implements the interface " );
16         p.addLn( resInterface.getName() + "[]." );
```

```
17      }
18      p.add( "public class " + joinClassName );
19
20      p.indent();
21          // Generate the 'join' method.
22          p.add( "public static Object join( Object arg1, Object arg2 ) " );
23          p.addLn( "throws ClassCastException" );
24
25          …
26
27          // Generate the 'match' method.
28          p.add( "static boolean match( " + tuple1ClassName + " tuple1, " +
29                  tuple2ClassName + " tuple2 )" );
30
31          p.indent();
32              p.add( "return (" );
33
34              for (int i = 0; i < commonAttributes.length; i++ ) {
35
36                  // Get the name of a common attribute.
37                  String attributeName = commonAttributes[ i ].getName();
38
39                  if ( i > 0 ) p.add( " && " );
40
41                  // Add a predicate to match on that attribute.
42                  p.add( "tuple1." + attributeName + "() == tuple2." +
43                          attributeName + "()" );
44              }
45
46              // Allow for no common attributes, in which case result is cross product.
47              if (commonAttributes.length == 0) p.add( "true" );
48
49              p.add( ");" );
50          p.outdent();
51      p.outdent();
52
53      return p.getText();
54 }
```

*Figure 16. Details of a generator method*

Figure 17 shows as an example the code generated by the *generateJoin* method for a natural join of relations with types *Employee[]* and *Job[]*. The generated class *Temp1* contains the main method *join* and two auxiliary methods, *match* and *concatenate*. Lines 7-8 in *join* cast the input relations to their specific types. The vector *resultVector* is created to store the result tuples as they are constructed. The nested loop in lines 13-18 then iterates through all of the possible pairs of tuples, creating a result tuple from each matching pair using the *match* and *concatenate* methods. An array representing the result relation is then created, and the tuples from the vector assigned into it. The array creation refers to the second generated class *Temp2*, which is the class of the result tuples.

The method *match* takes two parameters, one of each input tuple type. The result is a boolean expression consisting of the *and*ing together of a number of sub-expressions, one for each of the relational attributes common to the two input types. Each of these expressions compares the corresponding attribute values in the two tuples for equality. In fact, in this example there is only one sub-expression since the only common attribute is *jobId*.

The method *concatenate* returns a result tuple formed by copying the appropriate attribute values from a given pair of input tuples.

```java
1   import java.util.*;
2
3   // This is a generated class to join Employees and Jobs.
4   public class Temp1 {
5       public static Object join( Object arg1, Object arg2 ) throws ClassCastException {
6
7           Employee[] rel1 = (Employee[]) arg1;
8           Job[] rel2 = (Job[]) arg2;
9           Vector resultVector = new Vector();
10          int size1 = rel1.length;
11          int size2 = rel2.length;
12
13          for (int i = 0; i < size1; i++) {
14              for (int j = 0; j < size2; j++) {
15                  if ( match( rel1[ i ], rel2[ j ] ) )
16                      resultVector.addElement( concatenate( rel1[ i ], rel2[ j ] ) );
17              }
18          }
19
20          int resSize = resultVector.size();
21          Temp2[] resultArray = new Temp2[ resSize ];
22
23          for (int k = 0; k < resSize; k++)
24              resultArray[ k ] = (Temp2) resultVector.elementAt( k );
25
26          return resultArray;
27      }
28
29      static boolean match( Employee tuple1, Job tuple2 ) {
30          return ( tuple1.jobId() == tuple2.jobId() ); }
31
32      static Temp2 concatenate( Employee tuple1, Job tuple2 ) {
33          return new Temp2( tuple2.post(), tuple2.duties(), tuple2.jobId(),
34                            tuple1.name(), tuple1.title(), tuple1.department() ); }
35  }
```

*Figure 17. Part of the code generated by linguistic reflective Natural Join for* Employee[] *and* Job[]

The generated code shown in Figure 17 contains some parts which are specific to the particular types for which it was generated—for example the casting of the input relations, and the bodies of the *match* and *concatenate* methods—and some which are common to all classes produced by the generator. The balance between these depends on exactly how the generator is written. In this example it was chosen to generate separate *match* and *concatenate* methods because it makes the generated code easier to understand. Alternatively the bodies of those methods could have been generated inline in the *join* method for greater execution efficiency—a valid concern since the source code may be generated once and executed many times.

# EVALUATION

### Overview of Experiments

To assess the utility of linguistic reflection, the reflective implementation of generic natural join described in the previous section is evaluated in comparison with three other implementation techniques. These other techniques are as follows:

**tailored**     uses a non-generic function which operates over one specific pair of relation types, thereby fixing the input types at compilation time. Since no attempt is made to write generic code, this serves as a base for comparison with the other techniques. Any differences measured between this and the other, generic, functions thus represent overheads due to genericity.

**interpretive**     uses a single general type for relations and a single generic function which operates over that relation representation. With this solution, all relations are represented as instances of the same class. This corresponds to conventional relational systems.

**core reflective**     uses a generic function which operates over relations represented as typed arrays of tuple objects, in the same style as the linguistic reflective solution. The generic join function works by inspecting the types of the input relations, its subsequent actions depending on the structure of those types. The result type is passed to the function as a parameter. This solution uses a similar algorithm to the linguistic reflective solution, but executes it directly rather than generating new code that performs it.

The four techniques are evaluated with respect to several qualitative and quantitative criteria. The qualitative measures are:
- the difficulty of implementation;
- the differing points at which failure may occur.

The quantitative measures are:
- the overall execution times for an example workload;
- the specific execution times for forming relation representations, generating, compiling and loading code where applicable;
- the relative code sizes; and
- the amortisation of generation and compilation costs over multiple executions for the same input types using persistent caching of intermediate results.

**Details of Implementations Used for Comparison**

As outlined above, the base case for comparison—the *tailored* implementation—is a non-generic natural join function, tailored to the particular pair of input types required. The relations are represented as arrays of the appropriate tuple types. For example, for a test on input types *Employee* and *Job*, code similar to that shown in Figure 17 is used, with the relations represented as *Employee[]* and *Job[]* arrays. The only difference is that the specific types are used directly in the method signature:

```
public static Temp2[] join( Employee[] rel1, Job[] rel2 ) …
```

In fact, in the experiments, the tailored implementations are obtained by recording the temporary source files generated by the linguistic reflective implementation and amending the method signatures.

The *interpretive* implementation involves defining a single representation that can be used for all relations. This is the class *Relation*, an outline of which is shown in Figure 18. This shows the parameters of the constructor: an array of strings containing the attribute names; an array of classes containing the attribute domains; and a two-dimensional array of objects containing the attribute values of the tuples. The natural join operation is performed by the static method *natJoin*, which takes as parameters two instances of *Relation* and returns another as the result.

```
1   public class Relation {
2   
3       public Relation( String[] attributeNames, Class[] attributeDomains,
4                        Object[][] tuples ) throws InvalidRelationException {…}
5   
6       public static Relation natJoin( Relation rel1, Relation rel2 ) {…}
7   }
```

*Figure 18. Outline of generic* Relation *class*

Since the *Relation* class is sufficiently general to represent relations of any type, a number of validity checks have to be performed by the constructor. These include checking that no attribute names are duplicated, that each tuple has the correct number of attribute values and so on. The costs of these checks are incurred only in the interpretive implementation, and are measured in the experiments. In the other implementations, the use of strongly typed array representations for the relations means that the checks are performed automatically by the Java system.

The *core reflective* implementation, like the tailored implementation, uses different types to represent relations of different types. All of the genericity lies in the *natJoin* method, the signature of which is shown in Figure 19. It takes two arrays representing the input relations, the expected result class, and returns the result as an array. Since it works for all valid input and output relation types, the arrays are typed as *Object*.

```
1   public class NatJoin {
2       public Object natJoin( Object rel1, Object rel2, Class resTupleClass )
3           throws InvalidJoinException {…}
4   }
```

*Figure 19. Signature of* Core Reflective *implementation*

The *natJoin* method in the core reflective implementation uses the Java core reflection facilities to examine the type structure of the input relations in the same way as the linguistic reflective implementation. The building up of the result relation is then driven by that discovered structure. One difference between the linguistic reflective and core reflective implementations is that while the former generates new code to build the result, the latter performs the actions represented by that code immediately. Another difference is that the linguistic reflective implementation synthesises the type of the output relation, by generating a new class definition. In contrast, the core reflective implementation relies on the result class being passed as a parameter since Java's core reflection facilities provide no mechanism for creating new classes dynamically.

**Measurements**

The measurements are performed using the persistent Java system PJama Release 0.3.4.6[§§] in the default configuration. The platform is a 110 MHz Sun SPARCstation-4 with 32 MB RAM and 68 MB swap space, running Solaris 2.5.1 in single user mode. The timings are obtained by inserting calls to the method *System.currentTimeMillis()* in the Java source code immediately before and after each operation to be measured. This method returns the current "wall clock" time with milli-second resolution. Except where otherwise noted, the linguistic reflective implementation uses the direct compiler invocation mechanism from Figure 9.

To identify effects due to objects being brought into the PJama object cache, both cold and warm measurements are made. The cold timings correspond to the start of a new PJama invocation, whereas the warm timings are made after executing a number of iterations. Each individual cold measurement is obtained using a new invocation of the PJama abstract machine. The warm measurements are obtained by running the join 20 times in succession on the same invocation of the abstract machine, discarding the first 10 measurements and recording the following 10. Each timing figure shown in the tables is the mean of 10 measurements unless noted otherwise, with the standard deviation shown in brackets.

For each algorithm, timings are obtained for the natural join running on the same three data sets, each consisting of a pair of relations. The data sets all contain the same relational types (*Employee* and *Job*), and thus the same number of attributes (4 and 3 respectively), but vary in the number of tuples. Table 1 shows the number of tuples in each relation, and their overall sizes.

|        | relation 1 |        | relation 2 |        | result relation |        |
| ------ | ---------- | ------ | ---------- | ------ | --------------- | ------ |
|        | tuples     | size   | tuples     | size   | tuples          | size   |
| join 1 | 100        | 3.2 KB | 15         | 360 B  | 153             | 7.3 KB |
| join 2 | 1000       | 32 KB  | 150        | 3.6 KB | 12612           | 605 KB |
| join 3 | 3000       | 96 KB  | 700        | 16.8 KB| 175433          | 8.4 MB |

*Table 1. Sizes of relations used*

The bar chart in Figure 20 shows an overview of the measured total mean cold execution times. The inset contains a magnified view of the bars for the *join 1* data set. For that data set—the smallest—the cost of the linguistic reflective solution without cacheing is around 10 times higher than any of the others, and 100 times higher than the tailored solution. For the largest data set, however, the two linguistic reflective solutions are cheaper than any of the other generic solutions, and fairly close to the tailored solution, which represents the lowest cost attainable using the naive, join algorithm.

---

[§§] Further information on PJama is available at:
    `http://www.sunlabs.com/research/forest/opj.main.html`

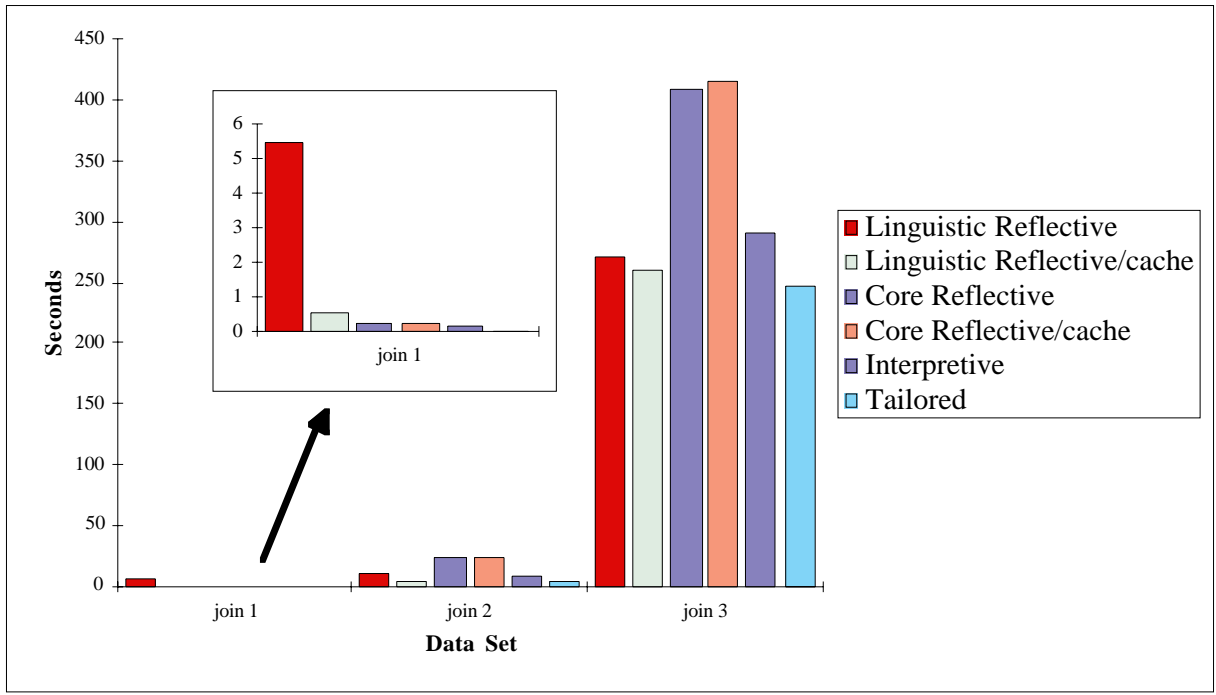

*Figure 20. Overview of join execution times*

Table 2 shows the timing figures in more detail. These show that there is only a small difference between cold and warm timings for the implementations other than the linguistic reflective ones. It was not possible to obtain warm figures for the largest data set due to problems with running out of main memory.

|        |      | tailored    | interpretive | core reflective | core reflective + cache | linguistic reflective | linguistic reflective + cache |
|--------|------|-------------|--------------|-----------------|-------------------------|-----------------------|-------------------------------|
| join 1 | cold | 0.05 (0.00) | 0.09 (0.00)  | 0.26 (0.00)     | 0.24 (0.01)             | 5.46 (0.01)           | 0.59 (1.70)                   |
|        | warm | 0.04 (0.00) | 0.08 (0.00)  | 0.23 (0.01)     | 0.24 (0.01)             | 3.74 (0.09)           | 0.04 (0.00)                   |
| join 2 | cold | 3.67 (0.03) | 8.23 (0.04)  | 24.37 (0.10)    | 24.75 (0.15)            | 11.04 (0.27)          | 4.48 (2.03)                   |
|        | warm | 3.56 (0.17) | 7.56 (0.30)  | 25.40 (0.26)    | 25.99 (2.20)            | 7.44 (0.26)           | 3.89 (0.16)                   |
| join 3 | cold | 246.9 (4.0) | 289.0 (43.5) | 407.4 (3.7)     | 415.0 (2.0)             | 269.1 (7.3)***        | 259.9 (9.1)                   |
|        | warm | N/A         | N/A          | N/A             | N/A                     | N/A                   | N/A                           |

*Table 2. Join execution times (seconds)*

For the linguistic reflective solutions there is a significant difference between cold and warm timings, particularly on the smaller data sets. In the case without cacheing this is probably due to the initial overhead of loading the compiler classes. Where persistent cacheing is used, the main difference is due to the fact that generation and compilation are only performed on the first execution for each data set. On subsequent executions the compiled join method is retrieved directly from the persistent store. Thus the set of cold timings contains only one for which compilation was necessary, and the warm set contains no such timings. Because of this the mean cold timings are somewhat artificial: the mean is dictated by the number of timings in the set, since only the first involves a generation and compilation overhead. If the size of the timings set was increased, the mean would approach that of the warm set as the cost of generation and compilation was amortised over a progressively larger number of executions.

The difference in cost between the two linguistic reflective solutions, while high for the small data set, becomes less as the size of the data set increases. This is because the costs of generation and compilation, which give rise to the difference between the two solutions, become less significant for larger relations in comparison to the costs of performing the join itself.

---

*** Average of only 7 iterations.

The interpretive solution also involves another cost not included in the figures in Table 2: that of checking relations for validity when they are formed. These checks are not needed in the other solutions since they use a more precisely typed representation for the relations, so that most of the necessary checks are performed statically by the compiler. Table 3 shows the measured times for constructing the relations: they are significant compared with the execution time for the small data set, but become less so as the data size increases.

|        |      | relation construction |
|--------|------|-----------------------|
| join 1 | cold | 0.04 (0.00)           |
|        | warm | 0.04 (0.00)           |
| join 2 | cold | 0.34 (0.00)           |
|        | warm | 0.39 (0.00)           |
| join 3 | cold | 1.1 (0.0)             |
|        | warm | N/A                   |

*Table 3. Relation construction times for interpretive implementation (seconds)*

Table 4 shows a breakdown of the costs involved in the linguistic reflective solution. In order to focus on those costs, it contains only figures for the non-cacheing implementation, so that generation and compilation is performed on every execution. This also represents the situation in which a linguistic reflective program is called only once for a particular set of input types. The right-hand side of the table contains corresponding figures obtained using the alternative mechanism for dynamic compiler invocation, that of issuing a shell command to invoke the compiler in a new operating system process.

As expected, the costs of generation, compiling and loading remain relatively constant across the data sets, since the same code is generated in each case, while the costs of performing the join operation increase with the number of tuples processed. The ratio between these costs varies from around 80:1 for the smallest data set, to 1:40 for the largest. The compilation and loading costs of the shell invocation mechanism are roughly twice those of the direct invocation. As before, this difference is more significant for the smaller data sets.

It is not clear why the generation costs increase as much as they do for the larger data sets, since the only parameters to the method calls being timed are the details of the relational classes, which remain the same. The PJama object cache will be fuller for the larger data sets, but it seems unlikely that the slow-down is due to objects being evicted from the cache, since the result relations have not yet been constructed at this stage. Even for the largest data set the input relations occupy only 112 KB.

|        |      | direct invocation |              |             | shell invocation |              |             |
|--------|------|-------------------|--------------|-------------|------------------|--------------|-------------|
|        |      | generation        | compile/load | join        | generation       | compile/load | join        |
| join 1 | cold | 0.19 (0.00)       | 5.20 (0.01)  | 0.07 (0.00) | 0.19 (0.01)      | 9.17 (0.17)  | 0.07 (0.00) |
|        | warm | 0.17 (0.00)       | 3.49 (0.04)  | 0.08 (0.09) | 0.12 (0.00)      | 8.94 (0.04)  | 0.05 (0.00) |
| join 2 | cold | 0.34 (0.00)       | 5.43 (0.01)  | 5.26 (0.27) | 0.35 (0.00)      | 9.19 (0.06)  | 4.02 (0.02) |
|        | warm | 0.23 (0.11)       | 3.41 (0.16)  | 3.80 (0.29) | 0.30 (0.01)      | 9.52 (0.14)  | 3.56 (0.05) |
| join 3 | cold | 0.6 (0.0)         | 6.4 (0.1)    | 262.2 (7.3)[†††] | 0.6 (0.0)    | 9.5 (0.2)    | 259.9 (2.1) |
|        | warm | N/A               | N/A          | N/A         | N/A              | N/A          | N/A         |

*Table 4. Breakdown of linguistic reflective times (seconds)*

Table 5 shows the number of lines of Java source code used in the various implementations. The tailored solution and the two linguistic reflective variants use the same representation for the relations, comprising the definitions of the classes *Employee* and *Job*. The core reflective variants require in addition the definition of the result tuple class *EmpJob*. The interpretive solution uses instead the generic class *Relation*.

The variations in code size for the natural join operation reflect the relative complexities. The tailored implementation is the simplest since it is non-generic, while the linguistic reflective implementation with cacheing is the most complex. The linguistic reflective implementations also produce 78 lines of generated code, which is identical to the tailored code except for the method signature, and two additional lines which cast the input relations to their specific array types.

---

[†††] Average of only 7 iterations.

|  | tailored | interpretive | core reflective | core reflective + cache | linguistic reflective | linguistic reflective + cache |
|---|---|---|---|---|---|---|
| representing relation | 37 | 61 | 46 | 46 | 37 | 37 |
| natural join | 76 | 169 | 413 | 435 | 538 | 560 |
| generated | N/A | N/A | N/A | N/A | 78 | 78 |

*Table 5. Code size (lines of Java source)*

**Summary of Measurements**

In summary, the timing measurements compare the execution costs of a linguistic reflective implementation of a generic operation with those of a non-generic solution, and with an interpretive and core reflective generic solution. The results obtained indicate that linguistic reflection is expensive where the amount of computation performed by the generated code is small. As the amount of computation increases the overheads of linguistic reflection become less significant. Using a persistent cache also allows those overheads to be amortised over multiple calls that use the same input types. Furthermore, for large data sets the linguistic reflective solution is cheaper than the alternative generic solutions.

**Qualitative Evaluation**

Table 6 shows the main points at which it is possible for dynamic failure to occur in the three generic implementations. Perhaps surprisingly, since the main motivation for using the reflective solution is to achieve stronger static typing of the relations, failure may occur later with those solutions than with the interpretive solution. Both reflective implementations may fail at join call time, if the relation parameters typed as *Object* do not have specific array types. Once the parameters have been checked no further application-specific errors can occur. In the case of the interpretive solution, run-time errors are only possible at the points that relation representations are formed. Once constructed successfully, two relations may be joined without risk of further failures.

|  | relation construction | join call |
|---|---|---|
| interpretive | X |  |
| core reflective |  | X |
| linguistic reflective |  | X |

*Table 6. Possible dynamic failure points*

As suggested by the code size figures given in Table 5, there is a spectrum of implementation complexity, with the interpretive solution being the simplest generic solution, increasing through the core reflective and linguistic reflective solutions. As well as requiring most code, the linguistic reflective implementation is significantly more difficult to write and understand than the others, since the underlying join algorithm is expressed at the meta-level rather than as normal code.

**Overall Evaluation**

The suitability of linguistic reflection for a given generic application is affected by a trade-off among execution speed, safety and ease of implementation. As illustrated for the natural join example, linguistic reflection allows more static type checking and thus fewer run-time errors in cases where the required genericity exceeds that provided by polymorphic type systems. Linguistic reflection also supports more efficient execution due to the elimination of some run-time checks, at larger workload sizes. The disadvantage is the greater implementation complexity.

# RELATED WORK

Reflection in general may be described as the ability of a programming system to change itself. A number of quite different reflective systems exist; they can be categorised according to the following criteria:

- behavioural or linguistic;
- untyped or strongly typed;
- compile-time or run-time.

The first choice is whether the reflection is behavioural or linguistic. Behavioural reflection is concerned with examining and altering aspects of the system's implementation. Linguistic (also called structural) reflection is restricted to examining existing meta-data, and generating and incorporating new source and type definitions. Another choice is whether or not the reflection operates in the context of a rigorously enforced type system. Finally, the reflection may occur at compile-time or at run-time. Figure 21 shows a classification of a number of reflective systems in this way.

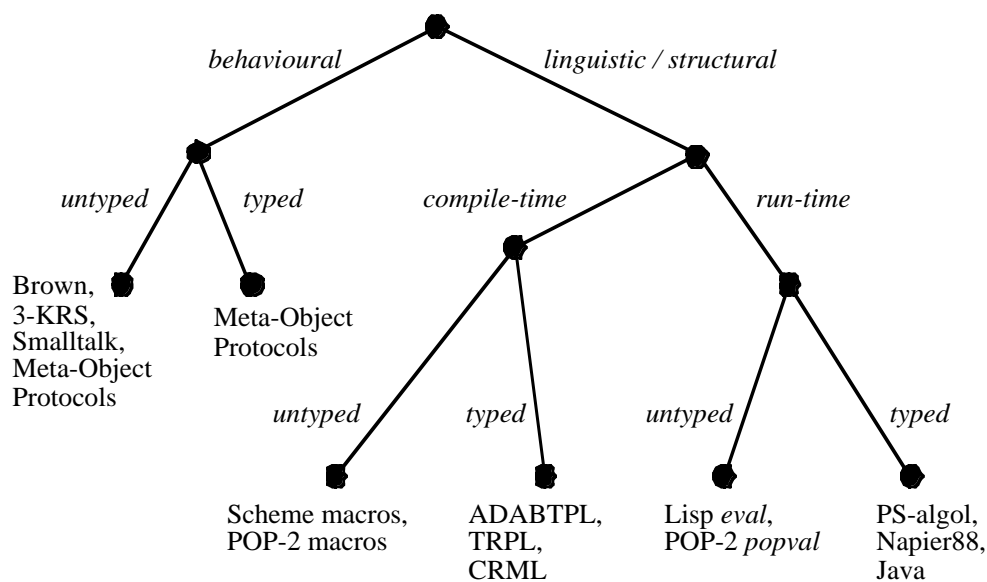

*Figure 21. Taxonomy of reflective systems*

**Behavioural Reflection**

Behavioural reflection allows introspection into aspects of a system's run-time behaviour. It is concerned not just with the programs that are running, but the process of their execution. Various aspects of the system that are normally implicit are made explicit to the programmer. For example, data structures representing the method dispatch mechanism or object storage layout may be made available. This provision of explicit meta-representations is known as reification.

The power of such systems is given by causally connecting the meta-representations back to the underlying structures and algorithms that they represent. A change to a meta-representation is reflected back in a corresponding change in the system behaviour. Thus fundamental aspects of the system may be altered dynamically under program control. Applications include tuning for efficiency on the basis of dynamic conditions, and as a software structuring technique for "Open Implementation"[14].

Behavioural reflection is thus different from linguistic / structural reflection in that it allows parts of the underlying system to be changed by executing programs, whereas linguistic reflection allows only the addition of new program fragments to an unchanged execution engine. One example of behavioural reflection in untyped systems is Brown[15], a variant of Lisp[16] that allows reflective functions to manipulate interpreter data structures. Another is 3-KRS[17], an object-oriented language in which every object has a corresponding meta-object containing information about the implementation and interpretation of the base object. By sending messages to its meta-object an object can alter itself indirectly. Smalltalk also provides reflective facilities[18]. Yet another approach is the Meta-Object Protocol (MOP) in CLOS[4]. More recently, MOPs have been developed for more strongly typed systems such as C++[19, 20].

**Linguistic Reflection**

Linguistic reflective systems are categorised as to whether the reflection and program generation occurs during compilation or at run-time. Compile-time reflection in an untyped language essentially gives a macro system[‡‡‡], such as in Scheme[21] or POP-2[22]. Strongly typed examples include ADABTPL[23], TRPL[1] and CRML[24]; these allow source code generating functions to be executed at compile-time. The results of these functions are incorporated into the program being processed, so at the end of compilation all reflective constructs have been compiled away leaving a conventional compiled program. Some applications of these languages have been in attaining high levels of genericity[23, 1], optimising implementations and validating specifications[25].

Finally, run-time linguistic reflection also appears in both untyped and typed systems. Untyped examples include the *eval* functions of Lisp[16] and the *popval* function of POP-2[22]. Strongly typed run-time linguistic reflection first appeared in PS-algol[7], and has also been used in Napier88[8] and Java—the subject of this paper. Applications include accommodation of system evolution[11, 26] and the partial automation of data model implementation[27, 28].

**Generators**

Another aspect in which the various linguistic reflective systems differ is the form of the generators which manipulate source code representations:

- Scheme and POP-2 programs may contain embedded macro calls. When such a call is encountered by the compiler the associated macro definition is executed. The macro may read lexical tokens from the current position in the compiler input stream, and generate new tokens which are inserted into the input stream for normal processing by the compiler.

- ADABTPL, TRPL and CRML also support macros (or functions in the case of CRML) which are executed during compilation. Within the body of a macro it is possible to access information derived by the compiler up to the point of compiling the macro call, for example the macro may discover the type of a particular variable. The result generated by the macro is an abstract syntax representation which is incorporated into the program being compiled, replacing the macro call.

- The *eval* function in Lisp and the *popval* function in POP-2 both dynamically evaluate source code representations, which may be generated by functions. In the case of Lisp the code representations are S-expressions, while in POP-2 they are lists of lexical tokens.

- In PS-algol and Java the source code representations are strings, which may be generated by procedures and methods respectively.

- Napier88 generators may produce either strings or hyper-program representations, which are program fragments that contain embedded direct links to persistent objects[29, 30]. Thus a value which exists at the time of generator execution may be bound directly into the generated code..[§§§]

**Relative Power of Reflective Mechanisms**

Behavioural reflection is a more powerful mechanism than linguistic reflection in terms of the kinds of changes that a reflective system may make to itself. It allows the underlying behaviour of the system to be changed, whereas linguistic reflection does not allow any changes that could not have been made by a programmer writing additional programs. The scope of possible alterations to system behaviour in a behaviourally reflective system depend on which of the implementation structures are made available for modification.

Because of its power and associated potential for misuse, behavioural reflection is principally used in systems in which flexibility is a high priority, such as untyped and rapid prototyping systems. It tends to be considered unsuitable for incorporation into strongly typed languages, unless strictly limited, since a program might use it to circumvent restrictions needed to enforce type safety. Strongly typed linguistic reflection, in contrast, may be introduced with complete safety, since no existing program code can be altered. Changes are restricted to the generation of new code, which must pass through normal compilation checks before it can be executed. Linguistic reflection in untyped languages is more powerful than in typed languages, in the sense that generated code fragments are subject to less checking before being executed. However, strongly typed systems have an advantage in this context in that a significant amount of descriptive information about the data being processed is automatically gathered during type checking. This information can then be used as the basis for reflective code generation.

---

[‡‡‡] There is no distinct boundary between macro and compile-time reflective systems. The latter can be thought of as macro systems with a relatively sophisticated execution environment.

[§§§] Further information on the Napier88 hyper-programming environment is available at:
   `http://www-ppg.dcs.st-and.ac.uk/`

Compile-time and run-time reflective systems differ in the time at which generation occurs, and thus in the environment available to the generator code. Generators in compile-time reflective systems have access to the compilation environment only, whereas run-time reflective generators have access to the run-time environment, allowing the code generated to be decided dynamically.

Most of the differences between various forms of code representation affect the style of the generator code, but not the power of the generators. The exception is the ability of Napier88 generators to manipulate hyper-program rather than simply text fragments. This allows objects in the generator execution environment to be bound directly into the generated source code. Generators in the other systems can only approximate to this by generating code that attempts to retrieve the objects from the environment in which the generated code is executed.

# FURTHER WORK

There are various possibilities for further work on Java-specific aspects of linguistic reflection, and more generally on providing better support for writing generators.

**Linguistic Reflection and Java**

One step to improve the integration of linguistic reflection with Java is to incorporate the dynamically callable compiler into the core reflection classes. This could be achieved by adding the class *Compiler* from Figure 7 to the package *java.lang.reflect*. Since this would require changing the core classes, a main memory based implementation of the compiler could be introduced at the same time.

Other improvements are possible with greater changes to the language implementation. Support for reflection over method source code would require alteration of compiled class formats, including *Class* instances and *.class* files, to include the source code. Thus a generator program could obtain the source code of a given class via the following method added to the class *java.lang.reflect.Method*:

```
public String getSource() throws IllegalAccessException
```

In some cases it would be undesirable for a class representation to include source code, so the choice of whether to include source code would be a compiler option. Attempted access to non-existent source code would result in an *IllegalAccessException*.

It is interesting to note that the availability of a dynamic compiler class subsumes the side-effecting methods in the core reflection package, such as *newInstance*, *set* and *invoke*. A program can instantiate classes, update object fields and invoke object methods by dynamically generating, compiling and executing code to do so. Pragmatically, it is desirable to retain the existing methods since they are simpler to use and more efficient than the equivalent linguistic reflective solutions. They do not however provide any extra expressive power.

The use of linguistic reflection in Java presents some language specific problems which merit further investigation. One difficulty is due to the name equivalence model for classes, in particular the fact that the subclass relationship must be stated explicitly rather than inferred from class structure. For example, an application might initially define the class *Person* and subclass *Student*. Later the need might arise for an intermediate class *UniversityMember*. Although such a class could be generated by linguistic reflection, it could not be introduced into the application since the class *Student* would still have the immediate superclass *Person*.

Another issue is the interaction between reflective programs and Java's access protection mechanism, which allows class members to be designated as public, protected or private. Protected and private members are not accessible outside the package in which they are defined. This fulfils both an abstraction role for modelling, and a security role to limit undesired access. In both cases the presence of protected and private members in a class limits the ability of reflective programs to inspect the structure of that class. In some cases this may be overly restrictive, for example an object browser which uses reflection to display the internal state of objects during debugging is unable to show protected and private members. Addressing these problems would require modification to the language.

Given that current implementations of Java are based on byte code interpretation, it would be possible for generators to generate byte codes directly, rather than source code representations. This would eliminate the compilation phase from linguistic reflection. While this may be attractive in terms of implementation efficiency, it would make generator programs considerably more complex. The generation of inconsistent byte code sequences could also compromise type safety.

**More General Issues**

Research is also needed into improving support for writing reflective programs, and generators in particular. This is less specific to Java, since understanding such programs appears to be hard in all existing linguistic reflective languages. The main difficulty is keeping track of both the generator code and the multiple possible generated code sequences. These are interspersed with one another and may appear in different forms. For example, generated code may be specified as abstract syntax expressions while the generator code appears as normal. Where string representations are used, constant expressions appear in the same form as normal code but it is then difficult for the programmer to distinguish between generator and generated code. Also since generator and generated code are evaluated in different environments, fragments of code which appear the same may have different meanings.

One approach to this problem is to provide tools which hide details of code manipulation, and help the programmer to focus either on the parts of the generated code which remain constant for all uses of the generator, or on the variable parts[31]. Another is to introduce a second level of generation, in which generators are themselves generated, allowing details that are common to all generators to be factored out[32].

Finally there is the issue of statically type-checking generators with respect to the code they generate, rather than just their own execution. This is a second order problem and appears to be unsolvable in general. Sheard has described one approach to restricting generators sufficiently that such checking can be achieved[33].

# CONCLUSIONS

This paper has focused on illustrating how strongly-typed run-time linguistic reflection may be provided in the compiled, strongly typed language Java, and how it may be used as a paradigm for program generation. We described the nature of linguistic reflection, how it can be integrated with a compiled strongly typed language, and how the facilities of Java JDK 1.1 can be augmented by a dynamically callable compiler to provide the basic facilities for linguistic reflection. As an example of the use of linguistic reflection, we described how it can be used to define a generic natural join function. Finally, we presented measurements which show that for some workloads, the technique supports generic implementations that are more efficient than other generic solutions.

Strongly typed run-time linguistic reflection allows programs to generate new programs safely. The addition of type-safety to linguistic reflection yields the following benefits:
- More information is available to the reflective computation, in the form of systematically required types. This information can be used to automatically adjust to implementation details and system evolution. Linguistic reflection without strong typing has little systematic information available about the structures involved in the computation.
- The type safety of all newly generated program fragments is checked before they are allowed to be executed. Such type discipline is highly advantageous in a programming environment in which the integrity of data must be supported.

It is, therefore, somewhat ironic that strong typing, which makes it difficult to integrate linguistic reflection with typed programming languages, also makes linguistic reflection effective as an amplifier of productivity. Conversely, linguistic reflection returns some of the expressibility to the language that was lost with the introduction of a type system (a major motivation for polymorphism). The importance of strongly typed linguistic reflection is that it provides a uniform mechanism for genericity and evolution that exceeds the capabilities of current non-reflective programming languages.

# ACKNOWLEDGEMENTS

This work is partially supported by the EPSRC through Grant GR/J 67611 "Delivering the Benefits of Persistence to System Construction" and by ESPRIT through Working Group EP22552 "PASTEL". We thank Dave Munro and the anonymous referees for their helpful suggestions.

# REFERENCES

1. T. Sheard, 'Automatic Generation and Use of Abstract Structure Operators' *ACM Transactions on Programming Languages and Systems*, **19**(4), 531-557 (1991).
2. D. Stemple, R. Morrison, G. N. C. Kirby and R. C. H. Connor, 'Integrating Reflection, Strong Typing and Static Checking', *Proceedings of 16th Australian Computer Science Conference*, Brisbane, Australia, 83-92 (1993).
3. D. Stemple et al., 'Type-Safe Linguistic Reflection: A Generator Technology' In **The FIDE Book**, M. P. Atkinson (ed), Springer-Verlag, To Appear (1998).
4. G. Kiczales, J. des Rivières and D. Bobrow, *The Art of the Metaobject Protocol*, MIT Press, Cambridge, Massachusetts 1991.
5. J. Kleinöder and M. Golm, 'MetaJava: An Efficient Run-Time Meta Architecture for Java™', *Proceedings of International Workshop on Object Orientation in Operating Systems*, Seattle, Washington (1996).
6. J. Malenfant, M. Jacques and F. Demers, 'A Tutorial on Behavioral Reflection and its Implementation', *Proceedings of Reflection 96*, San Francisco, 1-20 (1996).
7. PS-algol, 'PS-algol Reference Manual, 4th edition' Universities of Glasgow and St Andrews Technical Report PPRR-12-88 (1988).
8. R. Morrison, A. L. Brown, R. C. H. Connor, Q. I. Cutts, A. Dearle, G. N. C. Kirby and D. S. Munro, 'Napier88 Reference Manual (Release 2.2.1)' University of St Andrews (1996).
9. M. P. Atkinson, L. Daynès, M. J. Jordan, T. Printezis and S. Spence, 'An Orthogonally Persistent Java™' *SIGMOD Record*, **25**(4), 68-75 (1996).
10. A. Ohori, P. Buneman and V. Breazu-Tannen, 'Database Programming in Machiavelli - a Polymorphic Language with Static Type Inference' *ACM SIGMOD Record*, **18**(2), 46-57 (1989).
11. A. Dearle and A. L. Brown, 'Safe Browsing in a Strongly Typed Persistent Environment' *Computer Journal*, **31**(6), 540-544 (1988).
12. L. Cardelli and P. Wegner, 'On Understanding Types, Data Abstraction and Polymorphism' *ACM Computing Surveys*, **17**(4), 471-523 (1985).
13. L. Cardelli, 'Amber' In **Lecture Notes in Computer Science 242**, Springer-Verlag, 21-47 (1986).
14. G. Kiczales, 'Towards a New Model of Abstraction in Software Engineering', *Proceedings of IMSA '92 Workshop on Reflection and Metalevel Architectures* (1992).
15. D. P. Friedman and M. Wand, 'Reification: Reflection Without Metaphysics', *Proceedings of ACM Symposium on Lisp and Functional Programming*, 348-355 (1984).
16. J. McCarthy, P. W. Abrahams, D. J. Edwards, T. P. Hart and M. I. Levin, *The Lisp Programmers' Manual*, M.I.T. Press, Cambridge, Massachusetts 1962.
17. P. Maes, 'Concepts and Experiments in Computational Reflection', *Proceedings of OOPSLA'87*, Orlando, Florida, 147-155 (1987).
18. F. Rivard, 'Smalltalk: A Reflective Language', *Proceedings of Reflection 96*, San Francisco, 21-38 (1996).
19. B. Gowing and V. Cahill, 'Meta-Object Protocols for C++: The Iguana Approach', *Proceedings of Reflection 96*, San Francisco, 137-152 (1996).
20. Y. Ishikawa, A. Hori, M. Sato, M. Matsuda, J. Nolte, H. Tezuka, H. Konaka, M. Maeda and K. Kubota, 'Design and Implementation of Metalevel Architecture in C++ - MPC++ Approach', *Proceedings of Reflection 96*, San Francisco, 153-166 (1996).
21. J. Rees and W. Clinger, 'Revised Report on the Algorithmic Language Scheme' *ACM SIGPLAN Notices*, **21**(12), 37-43 (1986).
22. R. M. Burstall, J. S. Collins and R. J. Popplestone, *Programming in POP-2*, Edinburgh University Press, Edinburgh, Scotland 1971.
23. D. Stemple, L. Fegaras, T. Sheard and A. Socorro, 'Exceeding the Limits of Polymorphism in Database Programming Languages' In **Lecture Notes in Computer Science 416**, F. Bancilhon, C. Thanos and D. Tsichritzis (ed), Springer-Verlag, 269-285 (1990).
24. J. Hook and T. Sheard, 'A Semantics of Compile-time Reflection' Oregon Graduate Institute of Science & Technology (1994).
25. D. Stemple, T. Sheard and L. Fegaras, 'Linguistic Reflection: A Bridge from Programming to Database Languages', *Proceedings of 25th International Conference on Systems Sciences*, Hawaii, 844-855 (1992).
26. R. C. H. Connor, Q. I. Cutts, G. N. C. Kirby and R. Morrison, 'Using Persistence Technology to Control Schema Evolution', *Proceedings of 9th ACM Symposium on Applied Computing*, Phoenix, Arizona, 441-446 (1994).
27. R. L. Cooper, M. P. Atkinson, A. Dearle and D. Abderrahmane, 'Constructing Database Systems in a Persistent Environment', *Proceedings of 13th International Conference on Very Large Data Bases*, 117-125 (1987).
28. R. L. Cooper, 'On The Utilisation of Persistent Programming Environments' Ph.D. Thesis, University of Glasgow (1990).

29. G. N. C. Kirby, R. C. H. Connor, Q. I. Cutts, A. Dearle, A. M. Farkas and R. Morrison, 'Persistent Hyper-Programs' In **Persistent Object Systems**, A. Albano and R. Morrison (ed), Springer-Verlag, *Proceedings of the 5th International Workshop on Persistent Object Systems, San Miniato, Italy*, 86-106 (1992).
30. G. N. C. Kirby, R. C. H. Connor and R. Morrison, 'START: A Linguistic Reflection Tool Using Hyper-Program Technology' In **Persistent Object Systems, Tarascon 1994**, M. P. Atkinson, D. Maier and V. Benzaken (ed), Springer-Verlag, *Proceedings of the 6th International Workshop on Persistent Object Systems, Tarascon, France*, 355-373 (1994).
31. G. N. C. Kirby, 'Reflection and Hyper-Programming in Persistent Programming Systems' Ph.D. Thesis, University of St Andrews. Technical Report University of St Andrews Report CS/93/3 (1992).
32. R. Cooper and G. N. C. Kirby, 'Type-Safe Linguistic Run-time Reflection—A Practical Perspective' In **Persistent Object Systems, Tarascon 1994**, M. P. Atkinson, D. Maier and V. Benzaken (ed), Springer-Verlag, *Proceedings of the 6th International Workshop on Persistent Object Systems, Tarascon, France*, 332-354 (1994).
33. T. Sheard and N. Nelson, 'Type Safe Abstractions Using Program Generators' Oregon Graduate Institute Technical Report CS/E 95-013 (1995).